\documentclass[10pt,twocolumn,rmp,floats,amssymb,letterpaper]{revtex4}
\usepackage[dvips]{graphicx}
\usepackage{times}
\usepackage{mathptmx}

\bibpunct{}{}{,}{s}{}{}

\begin{document}

\title{Highly sensitive and broadband carbon nano\-tube radio-frequency single-electron transistor}

\author{S.~E.~S.~Andresen,$^1$ F.~Wu,$^2$ R.~Danneau,$^2$ D.~Gunnarsson,$^2$ and P.~J.~Hakonen\hspace{1.5pt}$^{2,}$\footnote{Electronic mail:
pjh@boojum.hut.fi}} \affiliation{\rmfamily\normalsize
$^1$\mbox{Niels Bohr Institute, Nano-Science Center, University of Copenhagen},\\
\mbox{}\hspace{0.75ex}\mbox{Universitetsparken 5, DK-2100 K\o benhavn \O, Denmark}\\
$^2$\mbox{Low Temperature Laboratory, Helsinki University of Technology,}\\
\mbox{}\hspace{0.75ex}\mbox{P.O.\ BOX 2200, FIN-02015 HUT, Finland}}

\date{April 16, 2008}

\begin{abstract}
\normalsize We have investigated radio-frequency single-electron
transistor operation of single-walled carbon nano\-tube quantum dots
in the strong tunneling regime. At a temperature of $4.2$ K and with
carrier frequency $754.2$ MHz, we reach a charge sensitivity of $2.3
\times 10^{-6}$ $e/\sqrt{\mathrm{Hz}}$ over a bandwidth of $85$ MHz.
Our results indicate a gain-bandwidth product of $3.7 \times
10^{13}$ $\mathrm{Hz}^{3/2}/e$, which is by one order of magnitude
better than for typical RF-SETs.
\end{abstract}

\maketitle

\section{Introduction}

The single-electron transistor (SET) is a highly sensitive
electrometer, conventionally based on sequential tunneling of
electrons in the Coulomb blockade regime. \cite{Likharev99}
Unfortunately, the $RC$-time due to cable capacitance ($C \sim 1$
nF) and device resistance ($R \sim 100 \mathrm{k}\Omega$) limits the
bandwidth to a few kHz. This is a major drawback for the direct
application of SETs as it limits the operation to a regime where
$1/f$-noise is strong, either due to background charge noise,
\cite{starmark99} or to variations in the tunneling resistance.
\cite{krupenin98}

The limitation on the frequency bandwidth can be bypassed using
microwave techniques for reading out the real part of the SET
impedance. In these so called RF-SETs, \cite{SchoelkopfScience99} an
$LC$-cir\-cuit is used to transform the high impedance of the SET to
a high-frequency setup with a characteristic impedance of $Z_0=50$
$\Omega$. The transformation is frequency selective, and a good
match is obtained only over a frequency range $f_0\pm f_0/Q$, where
$f_0=1/(2\pi\sqrt{LC})$ is the resonant frequency of the matching
circuit, and $Q=\sqrt{L/C}/Z_0$ under fully matched conditions.
Typical charge sensitivities of RF-SETs amount to $10^{-5}$
$e/\sqrt{\mathrm{Hz}}$ over a signal bandwidth of $20$ MHz.
\cite{SchoelkopfScience99}

In order to improve the performance of RF-SETs, the operating regime
has to be brought from sequential tunneling to co-tunneling, i.e.,
from the weak to the strong tunneling regime. \cite{Averin00} The
energy sensitivity including back action noise has been estimated to
be $\sim 2\hbar$ in the sequential tunneling regime.
\cite{DevoretSchoelkopfNature2000} In the co-tunneling regime, the
sensitivity is expected to approach $0.5 \hbar$. \cite{Averin00}
Another benefit of the strong tunneling regime is that a wider
bandwidth can be obtained as the $Q$-factor can be made smaller.
However, the effective Coulomb energy diminishes rapidly with
lowering resistance $R_{\mathrm{T}}$ of the tunnel barriers. This
behavior has been summarized for single junctions by Wang \emph{et
al.} \cite{WangEPL97} They find, e.g., that
$E_{\mathrm{C}}^{\mathrm{eff}}=0.3 E_{\mathrm{C}}$ for
$R_{\mathrm{T}}=3$ k$\Omega$. Provided that
$E_{\mathrm{C}}^{\mathrm{eff}} \gg k_{\mathrm{B}}T$, the strong
tunneling SETs are expected to operate well as RF-SETs.
\cite{Wahlgren} This regime of operation has been strived for by
Brenning \emph{et al.} \cite{BrenningJAP06} They managed to
fabricate Al/AlO$_x$ tunnel junction SETs with
$E_{\mathrm{C}}/k_{\mathrm{B}}=18$ K, and a total resistance of
$R_{\Sigma}=25$ k$\Omega$. For the charge sensitivity, they reached
$\delta q = 1.9\times 10^{-6}$ $e/\sqrt{\mathrm{Hz}}$ at 4.2 K,
which is the best liquid helium result obtained so far.

Single-walled carbon nanotubes (SWCNTs) provide an alternative
approach to the metallic SETs in the strong tunneling regime. The
first reports on single-electron charging effects in individual
tubes and bundles were published in 1997. \cite{tans97,bockrath97}
Charging energies of about $30$ meV were quickly observed.
\cite{nygard99} Since then, contacting techniques have greatly
improved, and impedances of $10$--$20$ k$\Omega$ can be rather
routinely obtained, e.g., using Pd contacts. \cite{Dai} SWCNTs with
large $E_{\mathrm{C}}$ are very promising for RF-SETs, especially
since their shot noise has been found to be well below the Schottky
value. \cite{FanSHOT}

In this article, we report on RF-SETs made from SW\-CNTs in the
strong tunneling regime. We find a charge sensitivity of $\delta q =
2.3\times 10^{-6}$ $e/\sqrt{\mathrm{Hz}}$ at $4.2$ K, which nearly
equals that obtained by Brenning \emph{et al.} \cite{BrenningJAP06}
Compared with previous car\-bon nano\-tube
RF-SETs,\cite{Lammi,BiercukPRB06,NotreDame} the improvement is by a
factor ranging from 7 to 200. In combination with a bandwidth of
\mbox{$85$ MHz}, our results represent a considerable improvement
for broadband charge sensing, e.g., for fast read-out of
single-electron devices such as quantum dots.

\section{Device fabrication and experimental setup}

Our SWCNTs are grown from patterned catalyst islands, following the
approach of Kong \emph{et al.} \cite{KongNat98,GroveNTT} RF-SET
operation necessitates the use of insulating substrates in order to
minimize the shunt capacitance. We use sapphire for lower losses and
charge noise compared to conventional Si/SiO$_2$. The chemical vapor
deposition (CVD) takes place in a ceramic tube furnace from a gas
mixture of Ar, H$_2$, and CH$_4$ at $\sim 900$ $^\circ\mathrm{C}$.
After growth, pairs of $25$/$15$ nm Ti/Au contacts, $0.3$ $\mu$m
apart, are defined between the catalyst islands by electron beam
litho\-graphy. A central top-gate, $0.1$ $\mu$m wide, is deposited
between the contacts. Under the actual $25$ nm Ti gate, an
insulating barrier is formed by five $2$ nm Al layers, each oxidized
$2$ min at atmosphere. Finally, we deposit a Cr/Au mask with $230$
$\mu$m bond-pads. The final device layout is illustrated in Fig.\
1(a).

\begin{figure}[tb]
\begin{center}
\includegraphics[scale=0.45]{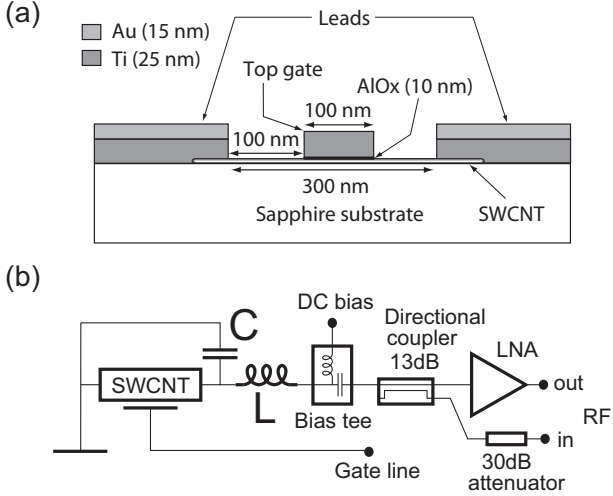}
\caption{(a) Layout side-view of the biasing leads and gate
electrode connected to the single-walled carbon nanotube. (b)
Experimental setup for RF-SET measurements, all immersed in liquid
helium for cooling to \mbox{$4.2$ K}.}
\end{center}
\end{figure}

A schematic of the low-temperature measurement setup is shown in
Fig.\ 1(b).\cite{WuLT24} The sample connects to an $LC$-circuit,
formed by the inductance \mbox{$L=150$ nH} of a surface mount
inductor and the bond-wires, and the parasitic capacitance
\mbox{$C=0.3$ pF} of the bond-pads. The circuit is connected to a
coplanar transmission line with a surface mount bias-tee that
couples the DC-bias and RF-signal. The top-gate is connected to a
separate coaxial line for high-bandwidth modulation. We use a
home-made low-noise amplifier with a frequency range of $600$--$950$
MHz. \cite{RoschierCryo04} The RF-input is coupled to the coaxial
line through a $13$ dB directional coupler and a $30$ dB attenuator
that reduces the noise from room temperature. The RF-output is
detected in a fashion dependent of the goal of the measurement: 1)
by a spectrum analyzer to investigate the carrier modulation
spectra, or 2) by mixer demodulation for homodyne detection at a
particular frequency.

\section{Results and discussion}

Fig.\ 2(a) shows the differential conductance
($dI/dV_{\mathrm{bias}}$) \mbox{versus} gate and bias voltages
($V_{\mathrm{G}}$, $V_{\mathrm{bias}}$). For $V_{\mathrm{bias}}>6$
mV, $dI/dV_{\mathrm{bias}}=3.0$--$3.5$ $e^2/h$, indicating a
high-quality SWCNT sample with highly transparent contacts. However,
there is a clear but smooth Coulomb modulation pattern at
$V_{\mathrm{bias}}=0$. As such, the sample behaves as a SET in the
strong tunneling regime. By tracing the Coulomb diamonds, we find
the addition energies $E_{\mathrm{add}}= 2E_{\mathrm{C}}+\Delta
E_{\mathrm{N}}$, with Coulomb energy
$E_{\mathrm{C}}=e^2/(2C_{\sum})$ and level spacing $\Delta
E_{\mathrm{N}}=E_{\mathrm{N}}-E_{\mathrm{N-1}}$. The values in Fig.\
2(b) cover the range $2.2$--$3.4$ meV, which sets an upper bound of
$E_{\mathrm{C}}/k_{\mathrm{B}}= 12.8$ K, corresponding to a total
capacitance of the SET island of $C_{\sum}\sim 73$ aF. From the gate
modulation and slopes of the diamonds in Fig.\ 2(a), we estimate
that the gate capacitance $C_{\mathrm{g}}\sim 2.9$ aF.

\begin{figure}[tb]
\begin{center}
\includegraphics[scale=0.75]{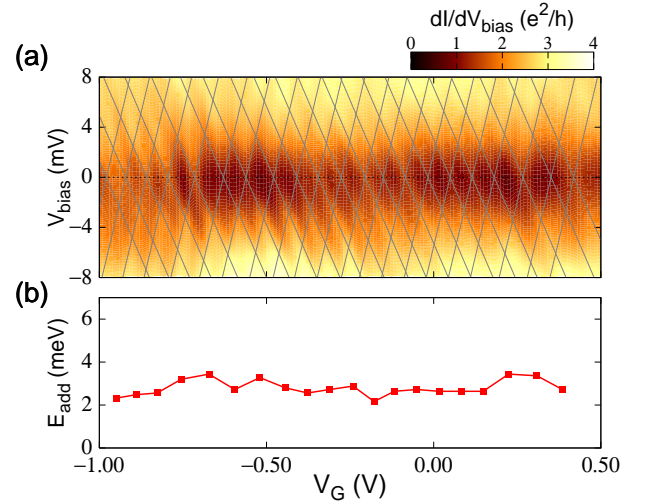}
\caption{(a) 2D-map of the differential conductance
$\mathrm{d}I/\mathrm{d}V_{\mathrm{bias}}$ in units of $e^2/h$ as a
function of gate voltage $V_{\mathrm{G}}$ and bias voltage
$V_{\mathrm{bias}}$. The lines mark the charge degeneracies. (b)
Values of the addition energy $E_{\mathrm{add}}$ deduced from (a) as
the degeneracy crossings in $V_{\mathrm{bias}}$.}
\end{center}
\end{figure}

For RF-SET operation, the optimum operating point was found by
searching for points of perfect matching, i.e., vanishing reflection
at the resonance frequency $f_0=754.2$ MHz. The signal was homodyne
detected by mixer, and the phase was tuned to be sensitive only to
the real part of the SET impedance. We found three points around
$V_{\mathrm{G}}=0.63$--$0.72$ V with maximum differential response,
coinciding with perfect matching (Fig.\ 3(a)). Using a spectrum
analyzer, the input carrier power was tuned to obtain maximum
signal-to-noise ratio of the sidebands at $f_0 \pm
f_{\mathrm{mod}}$, while keeping a small gate-charge modulation of
$q_{\mathrm{RMS}}=0.006\,e$ at \mbox{$f_{\mathrm{mod}}= 10$ MHz}
(Fig.\ 3(b)). The signal-to-noise ratio (SNR) of both sidebands
yields a charge sensitivity of $\delta q=2.3\times 10^{-6}$
$e/\sqrt{\mathrm{Hz}}$, corresponding to an uncoupled energy
sensitivity of $\epsilon=\delta q^2/(2C_{\Sigma})\sim 9\hbar$. The
frequency response was mapped out by repeating the sensitivity
measurement over a range of modulation frequencies of
$0.5$--\mbox{$150$ MHz} (Fig.\ 3(c)). We found a bandwidth of
$85\,\mathrm{MHz}$ and observed that $1/f$-noise only contributes
significantly below $1$--$2$ MHz.

\begin{figure}[tb]
\begin{center}
\includegraphics[scale=0.75]{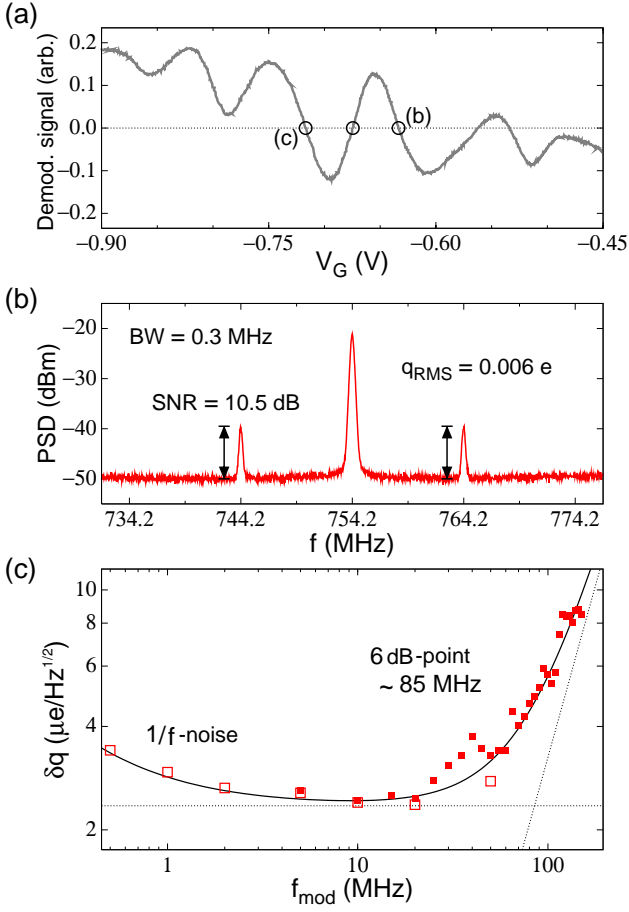}
\caption{(a) Demodulated signal averaged over $5000$ gate voltage
sweeps of each $2$ $\mu$s. Circles mark the points of perfect
matching used in (b) and (c). (b) Power spectrum ($0.3$ MHz spectral
resolution) under $754.2$ MHz carrier excitation ($-65$ dBm) and
$10$ MHz gate modulation of \mbox{$0.006\,e$}.
(c) Charge sensitivity versus modulation frequency
$f_{\mathrm{mod}}$, with/without a $100$ MHz low-pass filter on the
gate-line (open/filled symbols). The lines indicate the roll-off
limited by the bandwidth and $1/f$-noise at low frequency.}
\end{center}
\end{figure}

Optimization of RF-SET sensitivity has been treated in several
papers. \cite{korotokovAPL99,RoschierJAP04,TurinAPL03,TurinPRB04}
The ultimate shot noise limited sensitivity was found in Ref.\ 21 to
be
\begin{equation}
\delta q=2.65e\sqrt{\,\parbox[b][1.4em][b]{10.9ex}{$\displaystyle
R_{\Sigma}C_{\Sigma}\frac{k_{\mathrm{B}}T}{e^2/C_{\Sigma}}$}},
\end{equation}
which in our case ($R_{\Sigma}= 10$ k$\Omega$, $C_{\Sigma}= 73$ aF)
amounts to $0.9\times 10^{-6}$ $e/\sqrt{\mathrm{Hz}}$. When the
noise of the amplifier dominates the performance, the sensitivity
scales with the noise amplitude and the differential response of the
RF reflectance ($\partial\Gamma/\partial q$) as
\begin{equation}
\delta q =
\frac{\sqrt{\parbox[t][0.8em][b]{8.5ex}{$2\,k_{\mathrm{B}}T_{\mathrm{N}}\,
Z_0$}}}{\nu_{0}\times|\partial\Gamma/\partial q|},
\end{equation}
where $T_{\mathrm{N}}$ denotes the noise temperature of the
amplifier and $\nu_0$ the amplitude of the RF carrier
signal.
In our case, the amplitude over the SET is $\nu_{\mathrm{S}}=\nu_0
\sqrt{R_{\Sigma}/Z_0}\sim 2$ mV at perfect matching. In Ref.\ 23, an
approximation of Eq.\ (2) was found from the orthodox theory as
\begin{equation}
\begin{array}{l}
\displaystyle\frac{\delta q}{e}=1.46\times 10^{-6}\times
\bigg(\frac{k_{\mathrm{B}}T}{E_{\mathrm{C}}}\bigg)^{\!\!0.59}
\!\!E_{\mathrm{C}}(\mathrm{K})^{\:-1.01} \\
\displaystyle\qquad\quad\;\times
\bigg(\frac{R_{\Sigma}}{Z_{\mathrm{T}}}\bigg)^{\!\!0.91}
\!\!T_{\mathrm{N}}(\mathrm{K})^{\:0.5},
\end{array}
\end{equation}
with $E_{\mathrm{C}}$ and $T_{\mathrm{N}}$ in units of K, and
$Z_{\mathrm{T}}=\sqrt{L/C}$. As\-suming that heating does not play a
significant role, we estimate
from \mbox{Eq.\ (3)} that $\delta q=1.3\times10^{-6}$
$e/\sqrt{\mathrm{Hz}}$, taking $E_{\mathrm{C}}/k_{\mathrm{B}}=12.8$
K, $R_{\Sigma}=10$ k$\Omega$, $Z_{\mathrm{T}}=0.71$ k$\Omega$, and
$T_{\mathrm{N}}=4.2$ K. All of these estimates are based on the
classical orthodox theory for Coulomb block\-ade, \cite{Likharev99}
which is only valid in the weak tunneling regime. However, as
discussed above, we are in the strong tunneling regime. One way
around this issue is simply to adopt the effective charging energy
$E_{\mathrm{C}}^{\mathrm{eff}}=0.75 E_{\mathrm{C}}$, calculated by
Wang \emph{et al.}\cite{WangEPL97} Then the estimate for the charge
sensitivity becomes $\delta q=2.1\times10^{-6}$
$e/\sqrt{\mathrm{Hz}}$, which agrees well with the measured value.
In making our estimates, we have neglected the effects of quantum
level spacing (see discussion in Ref.\ 30).

\section{Conclusions}

We have shown that a SWCNT quantum dot in the strong tunneling
regime can be operated as an excellent RF-SET at liquid helium
temperature ($4.2$ K). We obtain a charge sensitivity of $\delta q =
2.3\times 10^{-6}$ $e/\sqrt{\mathrm{Hz}}$, which is nearly as good
as the best Al/AlO$_x$ results.\cite{BrenningJAP06} It represents an
enhancement of at least a factor of 7 compared with previous carbon
nanotube RF-SETs.\cite{Lammi,BiercukPRB06,NotreDame} Very recently,
comparable sensitivities were achieved with other types of SETs: 1)
InAs/InP he\-te\-ro\-structured nanowires,\cite{InAswire} and 2)
electrostatically defined Si quantum dots.\cite{AngusRFSET}

With a bandwidth of $85$ MHz, our gain-band\-width product amounts
to $3.7\times 10^{13}$ $\mathrm{Hz}^{3/2}/e$, where gain is defined
as $1/\delta q$. That is by more than one order of magnitude better
than for typical RF-SETs.\cite{SchoelkopfScience99} Therefore, SWCNT
based RF-SETs may have value in applications where high speed is
needed. One possible application is for charge sensing on quantum
dots in the context of quantum computing. Here, the coupling to a
nearby quantum dot structure, e.g., another nanotube, could be
achieved via an antenna gate as it was recently demonstrated with
SiGe nanowires.\cite{MarcusSiGe}

The best way to enhance the sensitivity further would be to lower
the temperature so the Coulomb modulation is fully developed. For
Al/AlO$_x$ junction SETs, the improvement at 40 mK was $\delta
q=0.9$ $e/\sqrt{\mathrm{Hz}}$ in the superconducting state and
$\delta q=1.0$ $e/\sqrt{\mathrm{Hz}}$ in the normal state.
\cite{BrenningJAP06} To improve the bandwidth, the only option is to
increase the resonance frequency of the $LC$-circuit. The limit set
by the Bode-Fano criterium states that the maximum achievable
bandwidth is $(2R_{\sum}C)^{-1}$, which in our case amounts to $\sim
170$ MHz.

\begin{acknowledgements}

We wish to acknowledge H.~I.~J\o rgensen, K.~Grove-Ras\-mussen,
T.~Heikkil\"a, M.~Paalanen, P.~E.~Lindelof, and B.~Pla\-cais for
fruitful discussions. This work was supported by the Academy of
Finland grant 213496 and by the EU under contract FP6-IST-021285-2.

\end{acknowledgements}


\begin{references}

\bibitem{Likharev99} \small See, \emph{e.g.}, K.\ K.\ Likharev, Proc.\ IEEE \textbf{87}, 606 (1999).

\bibitem{starmark99}
B.\ Starmark, T.\ Henning, T.\ Claeson, P.\ Delsing, and A.\ N.\
Korotkov, J.\ Appl.\ Phys.\ \textbf{86}, 2132 (1999).

\bibitem{krupenin98}
V.\ Krupenin, D.\ Presnov, M.\ Savvateev, H.\ Scherer, A.\ Zorin,
and J.\ Niemeyer, J.\ Appl.\ Phys.\ \textbf{84}, 3212 (1998).

\bibitem{SchoelkopfScience99}
R.\ J.\ Schoelkopf, P.\ Wahlgren, A.\ A.\ Kozhevnikov, P.\ Delsing,
and D.\ E.\ Prober, Science \textbf{280}, 1238 (1998).

\bibitem{Averin00} D.\ Averin, in \emph{Macroscopic Quantum Coherence and Quantum Computing}, edited by
D.\ V.\ Averin, B.\ Ruggiero, and P.\ Silvestrini (Kluwer, New York,
2001) pp. 399--408; arXiv:cond-mat/0010052.

\bibitem{DevoretSchoelkopfNature2000} M.\ H.\ Devoret, R.\ J.\ Schoelkopf, Nature \textbf{406}, 1039 (2000)

\bibitem{WangEPL97} X.\ Wang, R.\ Egger and H.\ Grabert, Europhys.\ Lett.\ \textbf{38}, 545 (1997).

\bibitem{Wahlgren}
P.\ Wahlgren, Ph.D.\ thesis, Chalmers University of Technology,
unpublished (1998).

\bibitem{BrenningJAP06} H.\ Brenning, S.\ Kafanov, T.\ Duty, S.\ Kubatkin and P.\ Delsing,
 J.\ Appl.\ Phys.\ \textbf{100}, 114321 (2006).

\bibitem{tans97}
S.\ J.\ Tans, M.\ H.\ Devoret, H.\ Dai, A.\ Thess, R.\ S.\ Smalley,
L.\ J.\ Geerlings, and C.\ Dekker, Nature (London) \textbf{386}, 474
(1997).

\bibitem{bockrath97}
M.\ Bockrath, D.\ H.\ Cobden, P.\ L.\ McEuen, N.\ G.\ Chopra, A.\
Zettl, A.\ Thess, and R.\ E.\ Smalley, Science \textbf{275}, 1922
(1997).

\bibitem{nygard99} J.\ Nygard, D.\ H.\ Cobden, M.\ Bockrath, P.\ L.\ McEuen, and P.\ E.\ Lindelof, Applied Physics A \textbf{69}, 297 (1999).

\bibitem{Dai} A.\ Javey, J.\ Guo, Q.\ Wang, M.\ Lundstrom, and H.\ Dai, Nature \textbf{424}, 654 (2003).

\bibitem{FanSHOT} F.\ Wu, P.\ Queipo, A.\ Nasibulin, T.\ Tsuneta, T.\ H.\ Wang, E.\ Kauppinen and P.\
J.\ Hakonen, Phys. Rev. Lett. \textbf{99}, 156803 (2007).

\bibitem{Lammi} L.\ Roschier, M.\ Sillanp\"a\"a, W.\ Taihong, M.\
Ahlskog, S.\ Iijima, and P.\ Hakonen, J.\ Low Temperature Phys.\
\textbf{136}, 465 (2004).

\bibitem{BiercukPRB06} M.\ J.\ Biercuk, D.\ J.\ Reilly, T.\ M.\ Buehler, V.\ C.\ Chan,
J.\ M.\ Chow, R.\ G.\ Clark, and C.\ M.\ Marcus, Phys.\ Rev.\ B
\textbf{73}, 201402 (2006).

\bibitem{NotreDame} Y.\ Tang, I.\ Amlani, A.\ O.\ Orlov, G.\ L.\
Snider, and P.\ J.\ Fay, Nanotech.\ \textbf{18}, 445203 (2007).

\bibitem{KongNat98} J.\ Kong, H.\ T.\ Soh, A.\ M.\ Cassell, C.\ F.\ Quate,
and H.\ Dai, Nature \textbf{395}, 878 (1998).

\bibitem{GroveNTT} K.\ Grove-Rasmussen, H.\ I.\ J\o rgensen, and P.\ E.\ Lindelof, Proceeding Int. Symp. on Mesoscopic Superconductivity and Spintronics 2006, NTT BRL, Atsugi, Japan, World Scientific Publishing (2007).

\bibitem{RoschierCryo04} L.\ Roschier and P.\ Hakonen, Cryogenics \textbf{44}, 783 (2004).

\bibitem{WuLT24} F.\ Wu, L.\ Roschier, T.\ Tsuneta, M.\ Paalanen, T.\ H.\ Wang, and
P.\ Hakonen, AIP Proc.\ \textbf{850}, 1482 (2006).

\bibitem{korotokovAPL99} A.\ N.\ Korotkov and M.\ A.\ Paalanen, Appl.\ Phys.\ Lett.\ \textbf{74}, 4052 (1999).

\bibitem{RoschierJAP04} L.\ Roschier, P.\ Hakonen, K.\ Bladh, P.\ Delsing, K.\ W.\ Lehnert, L.\ Spietz, and R.\ J.\ Schoelkopf, J.\ Appl.\ Phys.\ \textbf{95}, 1274 (2004).

\bibitem{TurinAPL03} V.\ O.\ Turin and A.\ N.\ Korotkov, App.\ Phys.\ Lett.\ \textbf{83}, 2898 (2003).

\bibitem{TurinPRB04} V.\ O.\ Turin and A.\ N.\ Korotkov, Phys.\ Rev.\ B \textbf{69}, 195310
(2004).

\bibitem{InAswire} H.\ A.\ Nilsson, T.\ Duty, S.\ Abay, C.\ Wilson,
J.\ B.\ Wagner, C.\ Thelander, P.\ Delsing, and L.\ Samuelson, Nano
Lett.\ \textbf{8}, 872 (2008).

\bibitem{AngusRFSET} S.\ J.\ Angus, A.\ J.\ Ferguson, A.\ S.\
Dzurak, and R.\ G.\ Clark, Appl.\ Phys.\ Lett.\ \textbf{92}, 112103
(2008).

\bibitem{MarcusSiGe} Y.\ Hu, H.\ O.\ H.\ Churchill, D.\ J.\ Reilly, J.\ Xiang, C.\ M.\ Lieber, and C.\ M.\
Marcus, Nature Nanotech.\ \textbf{2}, 622 (2007).

\bibitem{BeenakkerPRB1991}
C.\ W.\ J.\ Beenakker, Phys .\ Rev.\ B \textbf{44}, 1646 (1991).

\bibitem{quantumregime}

By studying the Coulomb diamond pattern around the optimum
sensitivity we deduce that $E_{\mathrm{C}} \sim \Delta
E_{\mathrm{N}}$ and $\Delta E_{\mathrm{N}} \sim 3 k_{\mathrm{B}} T$.
This suggests that we are in the quantum Coulomb blockade regime
(QCB), where $k_{\mathrm{B}} T \ll \Delta E_{\mathrm{N}} <
2E_{\mathrm{C}}$ and only few levels are involved in transport.
However, comparing the conductance peak line-shapes around optimum
sensitivity with theory \cite{BeenakkerPRB1991}, we find deviations
between the experimental data and QCB predic\-tions. From the
experimental data we get $G_{\mathrm{N}}=G/G_{\infty}\sim 0.45$ at
peak maximum and $dG_{\mathrm{N}}/dV_{\mathrm{G}} \sim 10$
$\mathrm{V}^{-1}$ as optimum transconductance. These values should
be compared to $G_{\mathrm{N}} \sim 0.75$ and
$dG_{\mathrm{N}}/dV_{\mathrm{G}} \sim 40$ $\mathrm{V}^{-1}$ as
predicted by theory. Also the predictions from the metallic Coulomb
blockade regime with $G_{\mathrm{N}} \sim 0.5$ and
$dG_{\mathrm{N}}/dV_{\mathrm{G}} \sim 20$ $\mathrm{V}^{-1}$ are
deviating, but the peak maximum is close to the measured. This
discrepancy is explained by the fact that we are in the strong
tunneling regime with $R_{\mathrm{Q}}/R_{\sum} \sim 2.5$ which
broadens the energy levels. Since we still have strong Coulomb
blockade we are not quite in the Breit-Wigner limit, but rather in
an intermediate regime $k_{\mathrm{B}} T < h\Gamma \sim \Delta
E_{\mathrm{N}} < 2E_{\mathrm{C}}$ which behaves closest to the
strong tunneling description.

\end{references}
\end{document}